\documentclass{ws-procs9x6-cpt22}
\begin{document}

\newcommand{\refeq}[1]{(\ref{#1})}
\def\etal {{\it et al.}}
%any other macros go here 

\title{Cosmology in symmetric teleparallel bumblebee gravity}

\author{J. E. G. Silva}

\address{Universidade Federal do Cariri (UFCA)\\
Av. Tenente Raimundo Rocha, \\ Cidade Universit\'{a}ria, Juazeiro do Norte, Cear\'{a}, CEP 63048-080, Brasil}

%\address{$^2$Group, Laboratory,\\
%City, State ZIP/Zone, Country}

%\author{On behalf of the LLG Collaboration}

\begin{abstract}
Symmetric teleparallel gravity (STG) is an alternative description of the gravitational field as an effect of the nonmetricity rather than the curvature of the spacetime. For an specific choice of the symmetric teleparallel lagrangian, the STG dynamics is equivalent to the general relativity (GR), and it is called Symmetric teleparallel equivalent to general relativity (STEGR). We propose a local Lorentz violating dynamics in the gravitational sector in the STEGR framework. The Lorentz violation is trigged by the vacuum expectation value (vev) of the well-known bumblebee field, whose vev couples to the nonmetricity tensor. The effects of such couplings on the cosmological evolution is investigated.
\end{abstract}

\bodymatter

\section{Introduction}

Since the pioneer work of Kostelecky and Samuel \cite{KS}, a great amount of investigations have been performed in search for signatures of Lorentz violation. In the gravitational sector of an effective framework called Standard Model Extension (SME), a self-interacting vector field called Bumblebee $B_\mu$ has a non-vanishing vacuum expectation value (vev) $b_\mu$ that triggers the Lorentz violation \cite{kostelecky,ks,ks2,maluf,seifert}. In fact, consider the following Einstein-Bumblebee action \cite{bertolami1,capelo,casana,malufbh,ovgun}
\begin{equation}
\label{einsteinbumblebeeaction}
S_g = \int{d^4x\sqrt{-g}\left(\frac{1}{2\kappa}R + \xi_1 B^\mu B^\nu R_{\mu\nu} -\frac{1}{4}B^{\mu\nu}B_{\mu\nu}-V(X)\right)},
\end{equation}
where $B_{\mu\nu}=2\partial_{[\mu} B_{\nu]}$ and $X=B^{\mu}B_\mu \pm b^2$. For $\xi_1 =0$ and $X=0$, the action in eq.(\ref{einsteinbumblebeeaction}) is nothing but the Einstein-Maxwell action. By setting $X=0$, i.e., frozen the bumblebee at its vev $b_\mu$, the non-minimal coupling $\xi_1 b^\mu b^\nu R_{\mu\nu}$ introduces the local Lorentz violation on the gravitational field. The effects of this coupling have been extensively studied in cosmology \cite{capelo}, black holes \cite{casana,malufbh}, wormholes \cite{ovgun}, among others. Extensions using the Kalb-Ramond field were also investigated \cite{kr1,kr2}. Finslerian spacetimes with a background vector were analysed in Ref.\cite{finslerkostelecky,finslersilva1,finslersilva,finslersilva2}.

Note that a coupling of type $b^\mu b^\nu b^\rho b^\sigma R_{\mu\nu\rho\sigma}$ vanishes due to the anti-symmetry of the Riemann tensor. Here, we propose other possible coupling in the framework of the symmetric teleparallel gravity.

%\begin{equation}
%S_B =\int{d^4x\sqrt{-g}\left(-\frac{1}{4}B^{\mu\nu}B_{\mu\nu}-V(X)\right)},
%\end{equation}

\section{Symmetric teleparallel Lorentz violations}

Let us consider a non-riemannian spacetime where the connection is not compatible with the metric. Instead, let us define the so-called non-metricity tensor $Q_{\mu\nu\rho}$ by \cite{fq1,fq2,fqpropagation1,fqpropagation2,fqcosmology2,fqblackhole}
\begin{equation}
Q_{\mu\nu\rho}=\nabla_\mu g_{\nu\rho}.
\end{equation}
A general symmetric connection can be split into
\begin{equation}
\Gamma^{\rho}_{\phantom{\alpha}\mu\nu}=\left\{^{\phantom{i} \rho}_{\mu\nu}\right\}  + L^{\rho}_{\phantom{\alpha}\mu\nu}\, ,
\end{equation}
where $\left\{^{\phantom{i} \rho}_{\mu\nu}\right\}$ are the usual Christoffel symbols and
\begin{equation}
L^{\rho}_{\phantom{\alpha}\mu\nu}  = \frac{1}{2} Q^{\rho}_{\phantom{\alpha}\mu\nu} - Q_{(\mu\phantom{\alpha}\nu)}^{\phantom{(\mu}\rho}\,,
\end{equation}
is called the disformation tensor. From the non-metricity tensor, we can define the so-called non-metricity vector by
\begin{equation}
Q^{\mu} =g^{\nu\rho} Q^{\mu}_{\ \nu\rho}\,, \quad \bar{Q}_{\mu}=\bar{Q}^{\nu}_{\phantom{\alpha}\nu\mu}\,.
\end{equation}
In addition, an invariant scalar, quadratic in the non-metricity components, can be defined by 
\begin{equation}
\mathcal{Q} =   -\frac{1}{4}Q_{\mu\nu\rho}Q^{\mu\nu\rho} +  \frac{1}{2}Q_{\rho\mu\nu}Q^{\mu\nu\rho} 
  +   \frac{1}{4}Q_{\mu} Q^{\mu}  
  - \frac{1}{2}Q_{\mu}\bar{Q}^{\mu}.
\end{equation}
The Ricci tensor for the connection $\Gamma^{\rho}_{\phantom{\alpha}\mu\nu}$ can be written as
\begin{eqnarray}
R_{\mu\nu}=\mathcal{R}_{\mu\nu}- L^{\alpha}_{\phantom{\gamma}\beta\mu}L^{\beta}_{\phantom{\gamma}\alpha\nu}-\frac{1}{2}Q_\alpha L^{\alpha}_{\phantom{\gamma}\mu\nu} + D_\alpha L^{\alpha}_{\mu\nu} + \frac{1}{2}D_\nu Q_\mu,
\end{eqnarray}
leading to the relation between the Ricci scalar and the non-metricity scalar
\begin{equation}
R   =  \mathcal{R}  + \mathcal{Q} +   \mathcal{D}_{\mu}( Q^{\mu} - \bar{Q}^{\mu} )\,.
\end{equation}
Assuming the teleparallel condition 
\begin{eqnarray}
R=0	& \Rightarrow & \mathcal{R}=-\mathcal{Q} -   \mathcal{D}_{\mu}( Q^{\mu} - \bar{Q}^{\mu} ).
\end{eqnarray}
Thus, the Einstein-Hilbert action leads to the same gravitational equations compared to those obtained from the variations of $-\sqrt{-g}\mathcal{Q}$. This is why this theory is called Symmetric teleparallel equivalent of general relativity (STEGR).
%\begin{equation}
%\int{d^4x \sqrt{-g}\mathcal{R}}=-\int{d^4x \sqrt{-g}\mathcal{Q}}.
%\end{equation}

Now let us consider Lorentz violating effects in the (STEGR) framework. The bumblebee vev $b_\mu$ can couple to the non-metricity vector leading to
\begin{equation}
S = \int d^4 x \sqrt{-g} \left(-\frac{1}{2\kappa}\mathcal{Q}+\lambda_1 b^{\mu}Q_\mu\right).
\end{equation}
The modified gravitational equation has the form
\begin{equation}
G_{\mu\nu} + \lambda_1 H_{\mu\nu}^{1} = \kappa T_{\mu\nu},
\end{equation}
where the tensor $H_{\mu\nu}^{1}$ is given by
\begin{equation}
H_{\mu\nu}^{1}=\frac{1}{\sqrt{-g}}\nabla_{\alpha}(\sqrt{-g}b^\alpha g_{\mu\nu})-\frac{1}{2}(b\cdot Q)g_{\mu\nu}-2 b^\alpha Q_{\alpha\mu\nu}.
\end{equation}

Let us investigate the effects of this Lorentz violating term in a cosmological model. Consider the flat Friedmann-Robertson-Walker spacetime
\begin{equation}
ds^2 = -dt^2 + a^{2}(t)\delta_{ij}dx^{i}dx^{j}.
\end{equation}
The bumblebee vev $b_\mu$ is chosen to be timelike in order to preserve the isotropy, i.e.,
\begin{equation}
b_\mu = (b_0, \vec{0}).
\end{equation}
The resulting Friedmann equations for a perfect fluid with equation of state $p=w\rho$ leads to
\begin{equation}
\label{friedmann}
\ddot{a}-\lambda_1 b \dot{a}=-\frac{\kappa}{6}(1+3w)\rho_0 a^{-3(1+w)+1}.
\end{equation}
Note that the linear Lorentz violating coupling yields a dissipative term $\lambda_1 b \dot{a}$. This term may has important influences during the inflation epoch, where $\dot{a}$ is large. Furthermore, this term may also contribute for the late-time accelerated expansion, leading to a de Sitter solution. These and other additional discussion open a new perspective on gravitational Lorentz violation and they will be properly analysed in a forthcoming work.

\end{document}